\documentclass[10pt,aps,prb,twocolumn,amsmath,amssymb,superscriptaddress]{revtex4-1}
\usepackage{amsmath}
\usepackage{graphicx}
\usepackage{color}
\usepackage{multirow}
\usepackage[export]{adjustbox}
\usepackage[caption = false]{subfig}
\usepackage{dcolumn}

\def\beq{\begin{eqnarray}}
\def\eeq{\end{eqnarray}}
\def\beqq{\begin{eqnarray*} \color{blue} }
\def\eeqq{\end{eqnarray*}}

\def\Nd{{N_d}}
\def\Nddiff{{N_{d}^{\mathrm{uniq}}}}

\begin{document}

\title{Excited states using semistochastic heat-bath configuration interaction}
\author{Adam A. Holmes}
\email{adamaholmes@gmail.com}
\affiliation{Department of Chemistry and Biochemistry, University of Colorado Boulder, Boulder, CO 80302, USA}
\affiliation{Laboratory of Atomic and Solid State Physics, Cornell University, Ithaca, New York 14853, USA}
\author{C. J. Umrigar}
\affiliation{Laboratory of Atomic and Solid State Physics, Cornell University, Ithaca, New York 14853, USA}
\author{Sandeep Sharma}
\email{sanshar@gmail.com}
\affiliation{Department of Chemistry and Biochemistry, University of Colorado Boulder, Boulder, CO 80302, USA}
\begin{abstract}
We extend our recently-developed heat-bath configuration interaction (HCI) algorithm, and our semistochastic algorithm for
performing multireference perturbation theory,
to the calculation of excited-state wavefunctions and energies.
We employ time-reversal symmetry, which reduces the memory requirements by more than a factor of two.
An extrapolation technique is introduced to reliably
extrapolate HCI energies to the Full CI limit.
The resulting algorithm is used to compute the twelve lowest-lying potential energy surfaces of the carbon dimer using the cc-pV5Z basis set, with an estimated error in energy of 30-50 $\mu$Ha
compared to Full CI. The excitation energies obtained using our algorithm have a mean absolute deviation of 0.02 eV compared to experimental values.
We also calculate the complete active-space (CAS) energies of the S0, S1, and T0 states of tetracene,
which are of relevance to singlet fission,
by fully correlating active spaces as large as 18 electrons in 36 orbitals. 
\end{abstract}
\maketitle

\section{Introduction}
The accurate \emph{ab initio} calculation of low-energy excited states is of great importance in many fields, including spectroscopy and
solar energy.
Unfortunately, excited-state calculations are complicated by the fact that they often exhibit strong multireference character,
and a simple and accurate variational \emph{ansatz} does not exist for them.
As a result, the commonly-used quantum chemical techniques such as density functional theory~\cite{hohenberg1964inhomogeneous,kohn1965self,parr1994density} and coupled cluster theory~\cite{coester1958bound,vcivzek1966correlation,cizek1980coupled,purvis1982full} become unreliable. Even when the excited states are qualitatively well-described by a single determinant, single reference methods such equation of motion coupled cluster singles and doubles~\cite{geertsen1989equation,stanton1993equation,nooijen1997new} (EOM-CCSD) are unable to describe states that have double-excitation character. 

Complete active space self-consistent field~\cite{roos1980complete1,roos1980complete2,siegbahn1981complete,roos2007complete} (CASSCF) and its extensions,
including multireference configuration interaction~\cite{werner1982self,siegbahn1983externally} (MRCI), complete active space perturbation theory~\cite{andersson1990second,finley1998multi} (CASPT2), n-electron valence perturbation theory~\cite{angeli2001introduction,angeli2001n,angeli2002n} (NEVPT2), variants of multi-reference coupled cluster~\cite{jeziorski1981coupled,laidig1984multi,lindgren1985linked} (MRCC), etc.,
are presently the most reliable methods for dealing with such problems. The shortcoming of these methods lies in their inability to treat problems that require more than 18 electrons and 18 orbitals in their active space, because of the need for performing exact diagonalization. To overcome this limitation, other methods such as restricted~\cite{malmqvist1990restricted,celani2000multireference} and generalized~\cite{ma2011generalized} active space (RASSCF/GASSCF) methods further subdivide the active space orbitals and put restrictions on their occupation pattern.
However, these methods quickly become unaffordable as one relaxes these restrictions to calculate exact active-space energies. Methods such as the density matrix renormalization group~\cite{white1992density,white1993density,white1999ab,schollwock2005density,chan2011density} (DMRG) algorithm,  Full Configuration Interaction Quantum Monte Carlo~\cite{Booth2009,Cleland2010} (FCIQMC) and its semistochastic improvement~\cite{PetHolChaNigUmr-PRL-12}, when used as active-space solvers, are able to go well beyond the restriction imposed by CASSCF and represent a significant advance.
They are nevertheless exponentially scaling, with the exception of DMRG for systems with a linear topology.
Although other approaches such as variational Monte Carlo~\cite{zhao2016efficient,mussard2017time,robinson2017excitation}, various flavors of projector Monte Carlo~\cite{schautz2004optimized,purwanto2009excited,guareschi2013ground,blunt2017density} 
and reduced density matrix based methods~\cite{NakEhaNak-JCP-02,mazziotti2006quantum,Val-ACP-07} 
have shown promise, they have not yet become widely used in quantum chemistry.


In this paper, we present a new, efficient excited-state algorithm using our recently-developed Heat-bath Configuration Interaction~\cite{HolTubUmr-JCTC-16} (HCI) algorithm,
in conjunction with semistochastic perturbation theory~\cite{ShaHolUmr-JCTC-17}.
The HCI method is in the category of Selected Configuration Interaction followed by Perturbation Theory (SCI+PT) algorithms.
The first SCI+PT algorithm, called Configuration Interaction by Perturbatively Selecting Iteratively (CIPSI), was developed over four decades ago by Malrieu and
coworkers~\cite{HurMalRan-JCP-73,EvaDauMal-CP-83},
and extended earlier Selected CI algorithms that did not use perturbation theory~\cite{BenDav-PR-69,whitten1969configuration}.
Many variations of CIPSI have been developed over the years~\cite{BuePey-TCA-74,cimiraglia1985second,cimiraglia1987recent,Knowles89,Har-JCP-91,povill1992treating,SteWenWilWil-CPL-94,GarCasCabMal-CPL-95,WenSteWil-IJQC-96,Neese-JCP-03,NakEha-JCP-05,AbrShe-CPL-05,BytRue-CP-09,Rot-PRC-09,Eva-JCP-14,Kno-MP-15,SchEva-JCP-16,LiuHof-JCTC-16,zhang2016deterministic,scemama2016quantum,garniron2017hybrid,giner2017jeziorski}.
These methods perform a pruned breadth-first search to explore Slater determinant space and identify those determinants that contribute the most to the targeted ground and excited states.
This step is followed by perturbation theory that includes the contributions from the first-order interacting space using Epstein-Nesbet perturbation theory~\cite{Eps-PR-26,Nes-PRS-55}.

HCI distinguishes itself from other SCI+PT methods in two respects. First, it changes the selection criteria, thereby allowing it to explore only the most important determinants without ever having to consider unimportant determinants (see Section~\ref{hci}). Second, it performs the perturbation theory using a semistochastic algorithm that eliminates the severe memory bottleneck of having to store the large number of determinants in the first-order interacting space (although memory-efficient deterministic variants are also possible, we have so far found them to be computationally much more expensive than the semistochastic algorithm).
These two ingredients make it more efficient than other SCI+PT approaches.
It is worth mentioning that, similar to FCIQMC and DMRG, the cost of the method scales exponentially with the size of the Hilbert space; however, for a large family of molecules, the prefactor is much smaller than the other algorithms, making HCI orders of magnitude faster.

Here we show that HCI can be made more efficient by utilizing time-reversal symmetry and angular
momentum symmetry, which is the largest abelian subgroup of the full $D_{\infty h}$ point group. We also present a new method for extrapolation of HCI energies to the full configuration interaction (FCI) limit. In contrast to the first two publications~\cite{HolTubUmr-JCTC-16,ShaHolUmr-JCTC-17} describing HCI, in which
we either extrapolated with respect to the variational parameter $\epsilon_1$, or assumed that our calculations were converged,
here we extrapolate the energies with respect to the perturbation energy $\Delta E_2$. 
We find that the extrapolation with respect to $\Delta E$ is often nearly-linear and more reliable than the previous extrapolation procedure.

The outline of the paper is as follows. In Sections~\ref{hci} and~\ref{spt}, we set the stage by describing the salient features of the HCI algorithm and semistochastic perturbation theory.
Next, in section~\ref{exc_states}, we show how the HCI algorithm can be extended to calculate excited states. In section~\ref{extrap} we present an improved method for extrapolating unconverged HCI energies to the FCI energies. In section~\ref{further} we describe further improvements to the algorithm, including the incorporation of angular momentum symmetries and time-reversal symmetry. Finally, in section~\ref{results}, we apply our new excited-state algorithm to calculate the S0, S1 and T0 states of the tetracene molecule and to the potential energy surfaces of 12 states of the carbon dimer, comparing to values from the literature where available. We finish the paper with some concluding remarks in section~\ref{conclusion}.

\section{Heat-bath Configuration Interaction (HCI)}\label{hci}
\subsection{Overview}
HCI is an efficient SCI+PT algorithm, which can be broken down into the following steps:

\begin{enumerate}

\item  Variational stage

\begin{enumerate}
\item  Identify the most important Slater determinants \label{step1}
\item  Find a variational wavefunction and energy by computing the ground state within the space spanned by determinants found in step \ref{step1}\label{step2}
\end{enumerate}

\item  Perturbative stage

\begin{enumerate}
\item  Identify the most important perturbative corrections to the variational energy\label{step3}
\item  Sum the contributions found in step \ref{step3}\label{step4}
\end{enumerate}

\end{enumerate}

Step \ref{step1} is performed as an iterative process, which alternates between adding new determinants to the selected space and finding the lowest-energy wavefunction within the current selected space. HCI improves over other SCI+PT algorithms by improving the algorithm for selecting the important determinants in steps 1a and 2a,
and also by using a semistochastic algorithm for performing the summation in step 2b that eliminates the need for storing all the determinants that contribute to the perturbative correction. 

\subsection{Variational stage}

In HCI, the variational wavefunction at any iteration is given by $\left|\psi\right\rangle=\sum_i^\mathcal{V} c_i \left|D_i\right\rangle$, and the new determinants $D_a$ that are added to the variational space are those for which 
$\max_{i\in \mathcal{V}}\left|H_{ai}c_i\right|$
is sufficiently large.
This is accomplished by defining a user-specified variational parameter $\epsilon_1$, and
adding new determinants $D_a$ if
\begin{align}
\left|H_{ai}c_i\right|>\epsilon_1
\label{eq:hci_ground}
\end{align}
for any determinant $D_i$ already present in the variational space with a coefficient $c_i$, which is connected to $D_a$ by a Hamiltonian matrix element $H_{ai}$.

The reason this criterion is used is because it can be implemented efficiently without checking the vast majority of the determinants that don't meet the criterion (see below). The determinants chosen using this scheme are approximately those chosen by the CIPSI algorithm, which chooses determinants $D_a$ for which
\beq
\left|c_a^{(1)}\right|=\left|\frac{\sum_i H_{ai}c_i}{E_0-E_a}\right|\label{eq:cipsi_ground}
\eeq
is sufficiently large.

The determinants chosen by the two criteria are nearly the same because the terms in the numerator of Eq.~\ref{eq:cipsi_ground} span many orders of magnitude, so the sum is highly correlated with the largest-magnitude term in the sum (eq~\ref{eq:hci_ground}). The denominators of Eq.~\ref{eq:cipsi_ground} also vary with $D_a$, but to a much lesser extent, since the determinant energies $E_a$ are
much larger than the current variational energy $E_0$ for sufficiently large variational expansions.


\subsection{Perturbative stage}

In SCI+PT algorithms, the perturbative correction $\Delta E_2$ is typically computed using Epstein-Nesbet perturbation theory,
\beq
 \Delta E_{2}=\sum_{a} \frac{\left(\sum_{i}H_{ai} c_i\right)^2}{E_0 - E_a}.
\label{eq:PT0}
\eeq
In the original CIPSI algorithm, this expression is computed by evaluating and summing \emph{all} of the terms in the double sum. However, the vast majority of the terms in the sum are negligible, so this 
approach is not very efficient.
Various schemes for improving the efficiency have been implemented, including only exciting from
a rediagonalized list of the largest-weight determinants~\cite{EvaDauMal-CP-83}, and its efficient approximation using
diagrammatic perturbation theory~\cite{cimiraglia1987recent}.
However, this is both more complicated than necessary (requiring a double extrapolation with respect to the two
variational spaces to reach the Full CI limit) and is more computationally expensive than necessary since even
the largest weight determinants have many connections that make small contributions to the energy.

HCI therefore introduced a ``screened sum'',
\beq
 \Delta E_{2}\approx\sum_{a} \frac{\left(\sum_{i}^{\left(\epsilon_2\right)}H_{ai} c_i\right)^2}{E_0 - E_a}.
\label{eq:PTc}
\eeq
where, $\sum^{\left(\epsilon_2\right)} H_{ai}c_i$ includes only terms for which
$\left\vert H_{ai} c_i \right\vert > \epsilon_2$.
Note that the vast number of terms that do not meet this criterion are \emph{never evaluated}.
Even with this screening, the number of connected determinants can be sufficiently large to exceed computer memory.
This is addressed in Sec.~\ref{spt}.

\subsection{Heat-bath algorithm for acceleration of both stages}
The key to the efficiency of the heat-bath scheme is as follows. The vast majority of the Hamiltonian matrix elements correspond to double excitations, and their values do not depend on the determinants themselves but only on the four orbitals whose occupancies change during the double excitation. Therefore, before performing an HCI
run, for each pair of orbitals, the set of all double-excitation matrix elements obtained by exciting from that pair of orbitals is computed and stored, sorted
in decreasing order by magnitude, along with the corresponding pairs of orbitals the electrons would excite to. Once this is done, at any point in the HCI algorithm, from a given reference determinant, all double excitations whose
Hamiltonian matrix elements exceed a cutoff (either $\epsilon_1/\left|c_i\right|$ or $\epsilon_2/\left|c_i\right|$ for the variational and perturbative stages, respectively) can be generated efficiently, \emph{without having to iterate over all double
excitations}.
This is achieved by iterating over all pairs of occupied orbitals in the reference determinant, and
traversing the list of sorted double-excitation matrix elements for each until the cutoff is reached.

This screening algorithm is utilized in both steps 1a and 2a of the algorithm,
and is a significant reason why the HCI algorithm is faster than other selected CI algorithms which do not truncate the search for double excitations, or
skip over the large number of determinants making negligible contributions to the energy.

\section{Semistochastic perturbation theory}\label{spt}
The evaluation of Eq.~\ref{eq:PTc} with a low computational cost requires the simultaneous storage of all included terms, indexed by $a$, and
can easily exceed memory limitations for challenging problems.

To overcome this memory bottleneck, we introduced a \emph{semistochastic} evaluation of the PT sum,
in which the most important contributions (found in the same way as in the original HCI algorithm) are evaluated deterministically
and the rest are sampled stochastically~\cite{ShaHolUmr-JCTC-17}.
Here, an initial deterministic perturbative correction $E_2^D[\epsilon_2^{\mathrm{d}}]$ is calculated using a relatively loose threshold $\epsilon_2^{\mathrm{d}}$.
Then, the stochastic calculation is used to only evaluate the error in the deterministic calculation by calculating the two stochastic energies $E_2[\epsilon_2]$ and $E_2[\epsilon_2^{\mathrm{d}}]$ (the second-order perturbative energy calculated with $\epsilon_2$ and $\epsilon_2^{\mathrm{d}}$ respectively) for every sample. The total second-order energy is given by the expression
\begin{align}
E_2 = (E_2[\epsilon_2] - E_2[\epsilon_2^{\mathrm{d}}]) + E_2^D[\epsilon_2^{\mathrm{d}}].
\end{align}
Both $E_2[\epsilon_2]$ and $E_2[\epsilon_2^{\mathrm{d}}]$ are calculated using the same set of samples, and thus there is significant cancellation of stochastic error. Furthermore, because these two energies are calculated simultaneously, the incremental cost of performing this calculation, compared to a fully-stochastic summation, is extremely small. Clearly, in the limit that $\epsilon_2^{\mathrm{d}} = \epsilon_2$, the entire perturbative calculation becomes deterministic.

The stochastic piece of the semistochastic summation algorithm
is evaluated by sampling only $N_d$ variational
determinants at a time. Each variational determinant $D_i$ is sampled, with replacement, with probability
\beq
p_i=\frac{\left|c_i\right|}{\sum_j \left|c_j\right|},
\eeq
using the Alias method~\cite{walker1977efficient,kronmal1979alias},
which has previously been used to efficiently sample double excitations in determinant-space quantum Monte Carlo algorithms~\cite{HolChaUmr-JCTC-16}.
It has been shown~\cite{ShaHolUmr-JCTC-17} that an unbiased estimate of the second-order perturbative correction to the energy is given by the expression
\begin{widetext}
\begin{align}
 \Delta E_{2} = \frac{1}{\Nd(\Nd-1)}  \sum_{a} \frac{1}{E_0 - E_a} \left[\left(\sum_{i}^{\Nddiff} \frac{ w_i  c_i H_{ai}}{p_i}\right)^2  +\sum_{i}^{\Nddiff} \left(\frac{w_i(\Nd-1)}{p_i } - \frac{ w_i^2}{p_i^2}\right)c_i^2 H_{ai}^2\right],
\end{align}
\end{widetext}
where $w_i$ denotes the number of times determinant $D_i$ is sampled, and the summation is only over the $N_d^{\mathrm{uniq}}$ unique determinants in the sampled $N_d$ determinants. A minimum of a mere two determinants is sufficient to perform this calculations, thus completely eliminating the memory bottleneck; however, the statistical noise is greatly diminished if larger samples can be chosen.

For large systems, the stochastic part of the algorithm is essential, but, for small systems it is possible to get within 1 mHa of the FCI energy
using just the deterministic part of the algorithm.  This is demonstrated for the C$_2$ dimer in the cc-pV5Z basis set in Table~\ref{breakdown}.
The variational plus the deterministic parts of the algorithm yield energies for the $^1\Sigma_g^+$ ground and excited states accurate
to 1 mHa in just 15 minutes on a single computer node.

We note that very recently a different semistochastic perturbation theory has been proposed~\cite{garniron2017hybrid}  wherein the statistical error
decreases much faster than the inverse square root of the number of Monte Carlo samples.

\begin{table*}[htb]
\caption{\label{breakdown}
The three contributions to the HCI energy, for the ground and first excited $^1\Sigma_g^+$ states of the carbon dimer at
$r=1.24253$ \AA\ in the cc-pV5Z basis set. In these calculations, $\epsilon_1=1\times 10^{-4}$ Ha, $\epsilon_2=1\times 10^{-8}$ Ha,
and the automatically-chosen $\epsilon_2^d$ values were found to be $2.9\times 10^{-6}$ Ha and $3.3\times 10^{-6}$ Ha for the
ground and excited states, respectively.
Natural orbitals from a separate state-averaged variational HCI calculation (also with $\epsilon_1=1\times 10^{-4}$ Ha) were used.
The FCI energy was found by extrapolation of several HCI runs, as described in section~\ref{extrap}.
The last column reports the CPU time in seconds on a single node (consisting of two 14-core 2.4 GHz Intel ``Broadwell'' processors), once the natural orbitals are obtained.
}
\begin{tabular}{c|d|d|c}
\hline
Component & \rm E_0 (Ha) & \rm E_1 (Ha) & Time (min)\\
\hline
\hline
Variational energy & -75.80598 & -75.71573 & 10\\ 
Deterministic component of PT correction & -0.00214 & -0.00223 & 5\\
Stochastic component of PT correction & -0.000045(6) & -0.00015(1) & 50\\
\hline
Total HCI energy & -75.808159(6) & -75.71811(1) & 65\\
\hline
\hline
Extrapolated total HCI energy & -75.80787(3) & -75.7190(1) \\
\hline
\end{tabular}
\end{table*}

\section{Excited states in HCI}\label{exc_states}

When computing ground and excited states, all states are expanded in the same set of variational determinants,
\beq
\left|\psi_{s}\right\rangle=\sum_{i\in \mathcal{V}} c_i^{(s)} \left|D_i\right\rangle,
\eeq
where $s$ denotes the state. This is akin to performing state-average variational calculations, rather than state-specific ones where each state would have its own set of determinants.
In the excited-state algorithm, at each iteration of the variational stage, we add to the variational space $\mathcal{V}$ the union of the new determinants that are important for
each of the states. Thus, each iteration, new determinants $D_j$ are added to $\mathcal{V}$ if
\beq
\left|H_{ji}\right|\left(\max_{s\in {\rm states}} \left|c_i^{(s)}\right|\right)>\epsilon_1\label{exc}
\eeq
for at least one $D_i\in \mathcal{V}$. Eq.~\ref{exc} ensures that when several states are targeted, the variational space will include more determinants than when only the ground state is targeted, since there will be determinants relevant to all targeted states. Note that this formula is different from the one used in state-average CASSCF theory where the density matrix is averaged which is closer in spirit to taking the square root of sum of the squares of the coefficients. This distinction becomes important in the event of degeneracies among the targeted variational states. In such a situation rotations within
the degenerate subspaces are arbitrary, and the value of the maximum magnitude of the coefficients is not invariant to such rotations. However, the square root of the sum of squares of coefficients is invariant. For such a situation we recommend using the invariant criterion, for example, if the ground state is nondegenerate but the first two
excited states are degenerate, one should use
$\max\left(\left|c_i^{(0)}\right|,\sqrt{\left|c_i^{(1)}\right|^2+\left|c_i^{(2)}\right|^2}\right)$
in place of $\max\left(\left|c_i^{(0)}\right|,\left|c_i^{(1)}\right|,\left|c_i^{(2)}\right|\right)$.
In the applications considered in this paper, there are no exact degeneracies among the
targeted states, so we use the simple formula in Eq.~\ref{exc}.

\section{Extrapolation of HCI energies}~\label{extrap}
Apart from the generalization to excited states, the most important modification to HCI in this paper
is a new procedure for extrapolation of the HCI total energy to the FCI limit. The HCI energy is a function of two parameters: $\epsilon_1$, which controls the variational stage,
and $\epsilon_2$, which accelerates the perturbative energy calculation by screening out the
many tiny contributions. In the limit that $\epsilon_1$ goes to zero, the HCI energy equals
the FCI energy, and in the limit that $\epsilon_2$ goes to zero, the perturbative correction
is exactly equal to the Epstein-Nesbet perturbation correction. In the calculations in this paper,
we use a fixed $\epsilon_2=10^{-8}$ Ha, which is sufficiently small to give near exact PT energies, and perform runs at several different values of $\epsilon_1$.

\begin{figure}
\begin{center}
\includegraphics[width=0.5\textwidth]{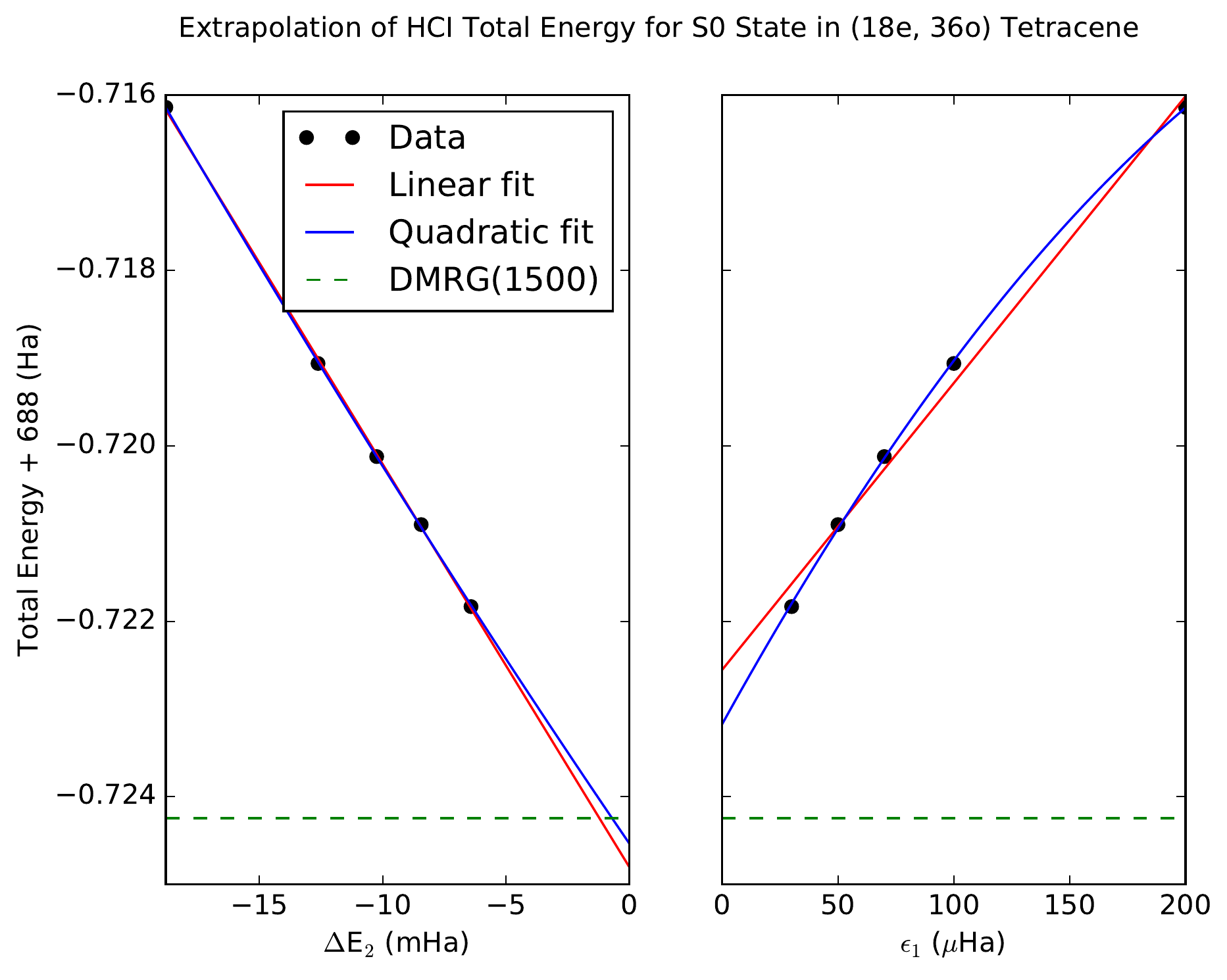}
\end{center}
\caption{Extrapolation of the HCI total energy to the FCI limit for the
lowest singlet state of tetracene in a DZ basis with an (18e, 36o) active space, using natural orbitals.
Previously, we obtained the FCI limit by extrapolating to $\epsilon_1=0$, using a rational polynomial
function of $\epsilon_1=0$, which requires several calculated values.
A linear or quadratic fit can yield an extrapolated value that is substantially in error, as shown
in the figure on the right.
In the present paper, we instead extrapolate to $\Delta E_2=0$, as shown
on the left. Not only is $\Delta E_2$ a more meaningful quantity than $\epsilon_1$, it also
enables a nearly-linear extrapolation of the total energy. The DMRG energy, used for comparison, is the variational
energy with bond dimension 1500 obtained by Hachmann et al~\cite{Hachmann2006}.
}\label{fig:extrap}
\end{figure}

In the original HCI paper~\cite{HolTubUmr-JCTC-16}, we extrapolated to the FCI limit by extrapolating the HCI energy
with respect to $\epsilon_1$.
However, this is often nonlinear with a curvature that increases as $\epsilon_1=0$ is approached.
Consequently, it is difficult to choose a function that provides a good fit to the computed energies.
Instead, in this paper, we extrapolate with respect to the perturbative correction to the energy.
In the limit that this perturbative correction is zero, both the variational and the total HCI energies equal the FCI energy.
In the limit that the extrapolation is linear, the variational and the total HCI energies extrapolate to precisely the same value.
As shown in figure \ref{fig:extrap}, this extrapolation is very close to linear.

\section{Further improvements to HCI}\label{further}

Since our most recent HCI paper~\cite{ShaHolUmr-JCTC-17}, in which we introduced the semistochastic algorithm for evaluating the HCI perturbative correction to the energy, we have improved the algorithm in several ways.
First, we have introduced the ability to employ angular momentum symmetry which is the largest abelian subgroup of the $D_{\infty h}$ point group
(for any but the smallest basis sets it has orbitals of a larger number of irreducible representations than for the $D_{2h}$ point group
that is commonly used for linear molecules).
Second, we have implemented time-reversal symmetry which can be used to perform separate calculations of the singlet and triplet
states, thereby reducing the Hilbert space of the problem by nearly a factor of two, and reducing the memory requirement of the Hamiltonian in the variational space by a factor of between two and four.

\subsection{Angular momentum symmetry}

For real orbitals the 2-electron integrals have 8-fold permutational symmetry.
Hence only slightly more than an eighth of the integrals need to be stored.
For linear molecules, the orbitals can be chosen to be eigenstates of the z-component of angular momentum, $\hat{L}_z$,
and the orbitals are complex.
In that case, there is only 4-fold permutational symmetry.
However, with the usual choice of phase, the integrals are real, and four of the eight
are zero since they violate $L_z$ conservation.  Hence it is still possible to store only an eighth
of the integrals, provided a check is performed to ensure $L_z$ conservation.
This enables us to use $L_z$ symmetry to reduce the storage required for the Hamiltonian without increasing the storage required for the integrals.

\subsection{Time-reversal symmetry}
The time-reversal operator exchanges the spin labels of the electrons.
States with $_z=0$ are symmetric/antisymmetric under time reversal if $S$ is even/odd.
Consequently the basis states can be chosen to be
symmetric or antisymmetric linear combinations of time-reversed pairs of Slater determinants.

Consider two spatial orbital configurations, $I$ and $J$. If a determinant is formed by assigning the
$\alpha$ electrons to $I$ and the $\beta$ electrons to $J$, i.e., $\left|I_\alpha J_\beta\right\rangle$,
then its time-reversed partner is $(-1)^{n_{\alpha}(n_{\beta}+1)}\left|J_\alpha I_\beta\right\rangle$, where $n_\alpha$ and $n_\beta$ are the number of alpha and beta electrons.
Note, $(-1)^{n_{\alpha}(n_{\beta}+1)}$ is always equal to 1 for a system containing an equal number of alpha and beta electrons, so we will ignore this phase from now on.
We choose to work in the basis of states $\left\{\left|S_{IJ}\right\rangle\right\}$, where 
\beq
\left|S_{IJ}\right\rangle = \begin{cases}
\left|I_\alpha J_\beta\right\rangle, & {\rm if } \quad I=J;\\
\frac{1}{\sqrt{2}} \left(\left|I_\alpha J_\beta\right\rangle+z\left|J_\alpha I_\beta\right\rangle\right), & {\rm if } \quad I\ne J,
\end{cases}
\eeq
where $z$ is the eigenvalue of the time-reversal operator,
which is either 1 for even $S$ states and -1 for odd $S$ states.
Note that basis states for which $I=J$ can occur only when $z=1$.

The matrix elements between pairs of these time-reversal symmetrized states are straightforwardly evaluated. For example, if $I=J$ and $K\ne L$,
\beq
\left\langle S_{IJ}\left| \hat{H} \right| S_{KL}\right\rangle
&=& \frac{1}{\sqrt{2}}\left\langle I_\alpha J_\beta \left| \hat{H} \right| K_\alpha L_\beta\right\rangle\nonumber\\
&&+ \frac{z}{\sqrt{2}}\left\langle I_\alpha J_\beta \left| \hat{H} \right| L_\alpha K_\beta\right\rangle,
\eeq
whereas if $I\ne J$ and $K\ne L$,
\beq
\left\langle S_{IJ}\left| \hat{H} \right| S_{KL}\right\rangle
&=& \frac{1}{2}\left\langle I_\alpha J_\beta \left| \hat{H} \right| K_\alpha L_\beta\right\rangle\nonumber\\
&&+ \frac{z}{2}\left\langle J_\alpha I_\beta \left| \hat{H} \right| K_\alpha L_\beta\right\rangle\nonumber\\
&&+ \frac{z}{2}\left\langle I_\alpha J_\beta \left| \hat{H} \right| L_\alpha K_\beta\right\rangle\nonumber\\
&&+ \frac{1}{2}\left\langle J_\alpha I_\beta \left| \hat{H} \right| L_\alpha K_\beta\right\rangle\nonumber\\
&=& \left\langle I_\alpha J_\beta \left| \hat{H} \right| K_\alpha L_\beta\right\rangle\nonumber\\
&&+ z\left\langle J_\alpha I_\beta \left| \hat{H} \right| K_\alpha L_\beta\right\rangle.
\eeq

We use time-reversal symmetry only for the variational stage. Upon completion of the variational stage, we
convert back to the determinant basis and perform Epstein-Nesbet perturbation theory in this basis.

Using time-reversal symmetrized states has two benefits. First, it shrinks the size of the Hilbert space, so that
a larger variational manifold can be treated with a given amount of memory. Second, it allows one to target different
symmetries separately. For example, if the ground state is a singlet and the first excited state is a triplet, then one can target the lowest triplet state
as a ground state and avoid using the excited-state algorithm.

\section{Results}\label{results}

We consider the excited states of two molecules, the carbon dimer and tetracene.

Despite its small size, the carbon dimer has strong multireference character even in its ground state, and has been
the focus of many experimental and theoretical studies~\cite{HydeWollaston1802, Wu1991, martin1992c2, Boggio-Pasqua2000, Danovich2004, UmrTouFilSorHen-PRL-07, Kokkin2007, Mahapatra2008, Varandas2008, TouUmr-JCP-08, purwanto2009excited, Booth2011, Shi2011, Su2011, Wang2011, Angeli2012, Brooke2013, Boschen2014, Wouters2014, Blunt2015, Krechkivska2015, Mayhall2015, Sharma2015, Krechkivska2016}.
Here we perform excited-calculations in Dunning's cc-pVQZ basis~\cite{dunning1989gaussian}
to compare to calculations from other methods in the literature. Then, we compute the twelve
lowest-lying potential energy surfaces in the larger cc-pV5Z basis, extrapolating to the Full CI limit.

The acenes are promising candidates for efficient solar conversion based on singlet fission~\cite{Smith2010, Zimmerman2010, Zimmerman2011, Lee2013, Smith}.
Here we compute the three lowest-lying states $-$ two singlets and one triplet $-$ in an active space
consisting of 18 pi electrons and either 18 or 36 pi orbitals in up to a cc-pVDZ basis.

All integrals used in these calculations were obtained using the PySCF quantum chemistry package~\cite{sun2017python}.

\subsection{Carbon dimer in cc-pVQZ basis}

In order to compare to DMRG and FCIQMC energies in the literature, we first computed the potential energy surfaces of the three lowest $^1\Sigma_g^+$ and two lowest $^5\Sigma_g$ states
in the cc-pVQZ basis with a frozen core.
These states were targeted by imposing $L_z=0$ ($\Sigma$ states) and using a basis of linear combinations of Slater determinants
which is symmetric under time-reversal symmetry (singlets and quintets).

\begin{table*}[htb]
\caption{\label{tab:qz}
Comparison of energies ($E+75$ in Ha) for the three lowest $^1\Sigma_g^+$ states of the frozen-core carbon dimer
in the cc-pVQZ basis set for three different methods.
The DMRG variational energies~\cite{Sharma2015} used a bond dimension of $M=4000$, and
were obtained by simultaneously targeting the three lowest states of $^1\Sigma_g^+$ symmetry.
The converged values for the DMRG variational energies of the ground state curve
(obtained by instead targeting only that single state) are slightly lower than those given here;
for example, the equilibrium energy is -75.80269(1) Ha, consistent with the extrapolated HCI total energy.
The HCI variational energies were obtained with $\epsilon_1=20$ $\mu$Ha,
and targeted the lowest five states of either $^1\Sigma_g^+$ or $^5\Sigma_g$ symmetry. They used state-averaged natural orbitals, which were obtained from an 
$\epsilon_1=50$ $\mu$Ha variational calculation of the lowest five states
in that symmetry sector.
Each HCI extrapolation was done using a linear extrapolation with respect to $\Delta E_2$ using two runs with
$\epsilon_1=50$ and 20 $\mu$Ha.
The uncertainties in the HCI total energies are 10-20 $\mu$Ha, and reflect both the stochastic
uncertainty in the individual points, as well as the
error in extrapolation to the FCI limit, taken to be 20\% of the difference in energy between the most accurate (smallest $\epsilon_1$) calculation and the extrapolated value.
The uncertainties in the FCIQMC total energies in Ref.~\onlinecite{Blunt2015}, range from 1 $\mu$Ha to 13 $\mu$Ha and
reflect only the stochastic noise; no attempt was made to extrapolate away the initiator bias.
}
\resizebox{\textwidth}{!}{%
\begin{tabular}{c|ccc|ccc|ccc|ccc}
\hline
\hline
$R/$\AA\ & \multicolumn{3}{c|}{DMRG Variational Energy} & \multicolumn{3}{c|}{FCIQMC Energy} & \multicolumn{3}{c|}{HCI Variational Energy} & \multicolumn{3}{c}{HCI Total Energy}\\
         & \multicolumn{3}{c|}{(Ref.~\onlinecite{Sharma2015})} & \multicolumn{3}{c|}{(Ref.~\onlinecite{Blunt2015})} & \multicolumn{3}{c|}{(this work)   } & \multicolumn{3}{c}{(this work)     }\\
\hline
1.0     & $-$ & $-$ & $-$ & -0.65570 & -0.48665 & -0.37654 & -0.65598 & -0.48688 & -0.37692 & -0.65620 & -0.48716 & -0.37725 \\
1.1     & -0.76124 & -0.62183 & -0.50228 & -0.76114 & -0.62170 & -0.50212 & -0.76103 & -0.62157 & -0.50196 & -0.76128 & -0.62186 & -0.50233\\
1.2     & -0.79920 & -0.69459 & -0.54490 & -0.79913 & -0.69450 & -0.54479 & -0.79901 & -0.69435 & -0.54461 & -0.79927 & -0.69465 & -0.54498\\
1.24253 & -0.80264 & -0.71208 & -0.54953 & -0.80258 & -0.71200 & -0.54942 & -0.80244 & -0.71182 & -0.54924 & -0.80271 & -0.71213 & -0.54961\\
1.3     & -0.79933 & -0.72633 & -0.54871 & -0.79927 & -0.72626 & -0.54861 & -0.79913 & -0.72607 & -0.54842 & -0.79939 & -0.72639 & -0.54881\\
1.4     & -0.77965 & -0.73267 & -0.53776 & -0.77961 & -0.73261 & -0.53766 & -0.77945 & -0.73240 & -0.53746 & -0.77973 & -0.73274 & -0.53789\\
1.6     & -0.72401 & -0.70487 & -0.51054 & -0.72395 & -0.70480 & -0.51047 & -0.72374 & -0.70457 & -0.51024 & -0.72410 & -0.70495 & -0.51072\\
1.8     & $-$ & $-$ & $-$ & -0.68056 & -0.65407 & -0.49639 & -0.68029 & -0.65389 & -0.49612 & -0.68071 & -0.65424 & -0.49661 \\ 
2.0     & -0.64552 & -0.61469 & -0.49290 & -0.64548 & -0.61470 & -0.49297 & -0.64522 & -0.61453 & -0.49269 & -0.64565 & -0.61486 & -0.49316\\
\hline
\hline
\end{tabular}
}
\end{table*}

The HCI variational and total energies are shown in Table~\ref{tab:qz}.
Note that even in a relatively large cc-pVQZ basis the variational energies are within 0.5 mHa of the converged total energies for
both ground and excited states.
The HCI total energies are lower than the (bond dimension 4000) DMRG and FCIQMC energies by 40-260 $\mu$Ha and 120-710 $\mu$Ha respectively.
For the ground state, DMRG energies that are in better agreement with HCI energies were obtained~\cite{Sharma2015}
by targeting just the ground state energy, e.g., the equilibrium energy is -75.80269(1) Ha, consistent with the extrapolated HCI total energy.
The discrepancy between the FCIQMC energies and the HCI total energies are likely due to the initiator bias in FCIQMC.

\subsection{Carbon dimer in cc-pV5Z basis}
We next computed the potential energy surfaces of the twelve lowest-lying states of the carbon dimer in the cc-pV5Z basis:
\beq
\Sigma_g &:& \quad X^1\Sigma_g^+,b^3\Sigma_g^-,B'^1\Sigma_g^+\nonumber\\
\Sigma_u &:& \quad c^3\Sigma_u^+,1^1\Sigma_u^-,2^3\Sigma_u^+\nonumber\\
\Pi_u &:& \quad a^3\Pi_u,A^1\Pi_u\nonumber\\
\Pi_g &:& \quad d^3\Pi_g,C^1\Pi_g\nonumber\\
\Delta_g &:& \quad B^1\Delta_g\nonumber\\
\Delta_u &:& \quad 1^3\Delta_u\nonumber
\eeq

\begin{table*}[htb]
\caption{The twelve lowest-lying potential energy curves ($E+75$ in Ha) of the frozen-core carbon dimer in the
cc-pV5Z basis, computed with HCI.
Linear extrapolations were performed on two runs with $\epsilon_1=100$ and 50 $\mu$Ha, and
the uncertainty was reduced by extrapolating with the average slope from all the linear extrapolations of points across a given potential energy surface.
The uncertainties in the HCI total energies are approximately 30-50 $\mu$Ha, and
include both stochastic error in individual points,
as well as error in extrapolation, taken to be 20\%
of the difference in energy between the most accurate calculation and the extrapolated value.
\label{table:c2_v5z}}
\resizebox{\textwidth}{!}{%
\begin{tabular}{c|c|c|c|c|c|c|c|c|c|c|c|c}
\hline
\hline
$R/$\AA\ & $X^1\Sigma_g^+$ & $a^3\Pi_u$ & $b^3\Sigma_g^-$ & $A^1\Pi_u$ & $c^3\Sigma_u^+$ & $B^1\Delta_g$ & $B'^1\Sigma_g^+$ & $d^3\Pi_g$ & $C^1\Pi_g$ & $1^1\Sigma_u^{+/-}$ & $1^3\Delta_u$ & $2^3\Sigma_u^{+/-}$\\
\hline

1.0  & -0.66252 & -0.58952 & -0.50414 & -0.55013 & -0.64546 & -0.46581 & -0.49405 & -0.54266 & -0.48560 & -0.47169 & -0.28627 & -0.31376 \\
1.1  & -0.76701 & -0.72504 & -0.66260 & -0.68697 & -0.73883 & -0.62800 & -0.62799 & -0.66169 & -0.60129 & -0.57169 & -0.44793 & -0.47545 \\
1.2  & -0.80461 & -0.78657 & -0.74165 & -0.74986 & -0.76505 & -0.71040 & -0.70037 & -0.70862 & -0.64496 & -0.60667 & -0.53547 & -0.56209 \\
1.3  & -0.80444 & -0.80488 & -0.77373 & -0.76951 & -0.75408 & -0.74553 & -0.73184 & -0.71485 & -0.64848 & -0.60560 & -0.57848 & -0.60277 \\
1.4  & -0.78460 & -0.79879 & -0.77864 & -0.76481 & -0.72483 & -0.75322 & -0.73799 & -0.70009 & -0.65680 & -0.58606 & -0.59644 & -0.61543 \\
1.5  & -0.75663 & -0.77988 & -0.76845 & -0.74736 & -0.69009 & -0.74566 & -0.72893 & -0.67759 & -0.66995 & -0.55851 & -0.60848 & -0.61240 \\
1.6  & -0.72895 & -0.75524 & -0.75062 & -0.72416 & -0.65898 & -0.73027 & -0.70975 & -0.65710 & -0.67153 & -0.62366 & -0.61900 & -0.61493 \\
1.7  & -0.70582 & -0.72897 & -0.72953 & -0.69951 & -0.63623 & -0.71147 & -0.68417 & -0.64216 & -0.66652 & -0.62777 & -0.62295 & -0.61793 \\
1.8  & -0.68550 & -0.70334 & -0.70780 & -0.67575 & -0.62333 & -0.69188 & -0.65860 & -0.63000 & -0.65797 & -0.62692 & -0.62214 & -0.61352 \\
1.95 & -0.65837 & -0.66874 & -0.67677 & -0.64456 & -0.61498 & -0.66407 & -0.62720 & -0.61373 & -0.64270 & -0.62083 & -0.61609 & -0.59905 \\
2.1  & -0.63561 & -0.64012 & -0.64963 & -0.62025 & -0.60665 & -0.64009 & -0.60572 & -0.60043 & -0.62760 & -0.61257 & -0.60767 & -0.58904 \\
2.4  & -0.60433 & -0.60182 & -0.60940 & -0.59236 & -0.59148 & -0.60643 & -0.58535 & -0.58525 & -0.60404 & -0.59723 & -0.59205 & -0.57999 \\

\hline
\hline
\end{tabular}
}
\end{table*}

Besides spatial symmetry, time-reversal symmetry was used to further reduce the size of the Hilbert space by targeting singlets (or quintets)
and triplets separately. Thus, for example, the three $\Sigma_g$ states were computed in two runs:
one which targeted the two lowest-energy singlets and one which target only the lowest energy triplet.

To accelerate convergence of the HCI total energy with respect to $\epsilon_1$, HCI natural orbitals
were used. Within each of the spatial symmetry sectors, the natural orbitals corresponding to the
state-averaged 1-RDM of the lowest variational states of interest were computed.
Thus, at each geometry, ten sets of natural orbitals were computed, one for each of the ten
spatial symmetry sectors.
After the natural orbitals were obtained, for each of the ten symmetry sectors, at each geometry,
at least two HCI runs were performed with those natural orbitals in order to enable
extrapolation to the FCI limit.


Each HCI calculation was performed starting from a single basis state
of the target irreducible representation, found automatically using the following algorithm.
First, estimate the global lowest-energy determinant by filling the orbitals with the lowest
one-body integrals; this is the current best guess for a good HCI starting state. Next, repeat the
following step until convergence: Replace the current HCI starting state with the basis state of the
target irreducible representation with lowest energy out of the set of states which are no more than
a double excitation away from the current HCI starting basis state.
%

Although this algorithm does not necessarily
result in either the lowest-energy basis state of the target symmetry sector, or the one with maximum
overlap with the Full CI ground state, it resulted in a good enough
starting point for the HCI runs in this paper.

\begin{figure*}[htb]
\begin{center}
\includegraphics[width=0.8\textwidth]{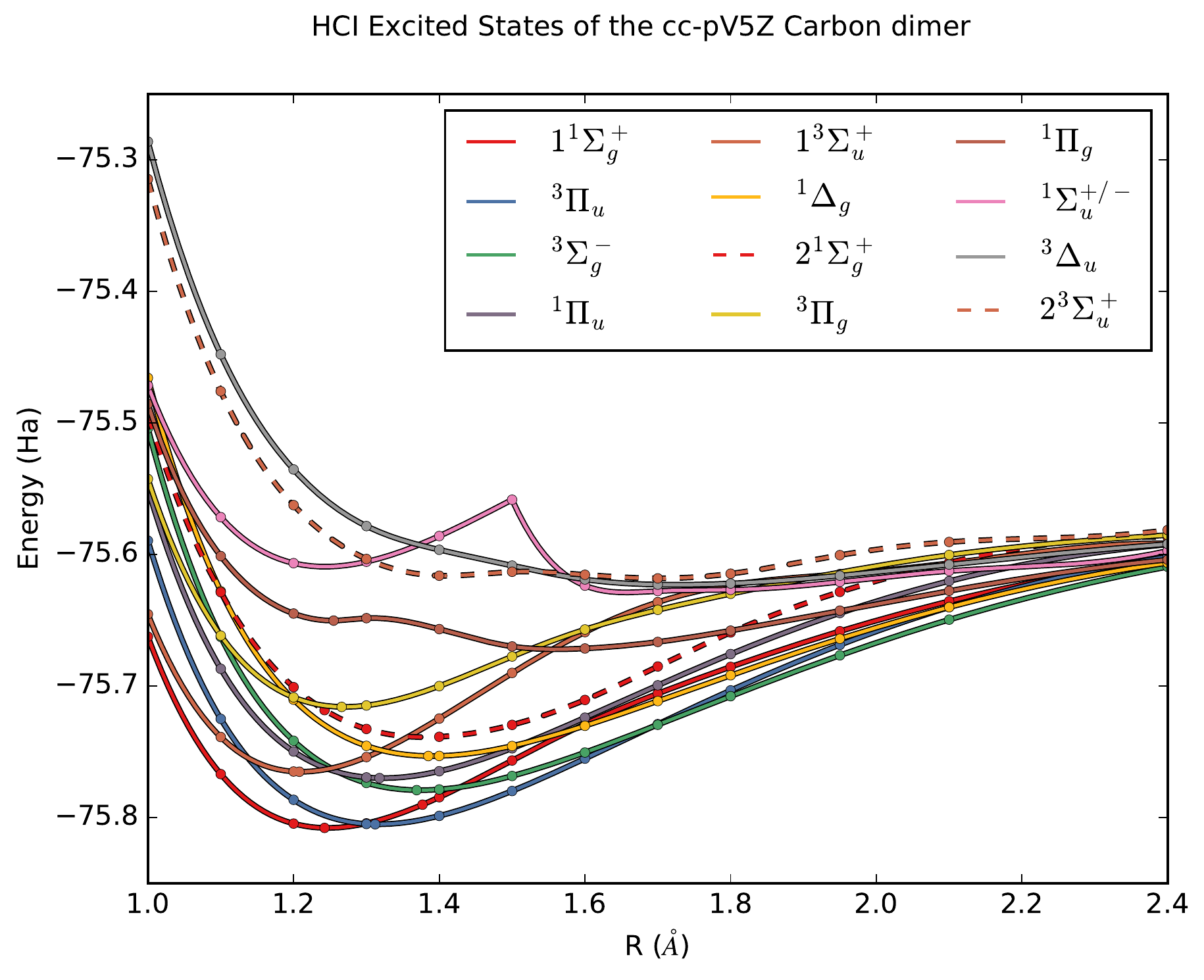}
\end{center}
\caption{
The twelve lowest-lying potential energy surfaces of the carbon dimer in the cc-pV5Z basis, computed using
HCI (E+75 in Ha). Time-reversal and $L_z$ symmetry were used, and cores were frozen. At each geometry, ten HCI runs were
performed, targeting either the one or two lowest states in each of the ten symmetry sectors.
The dotted lines correspond to the states that were computed using the excited state algorithm, while the
solid lines were computed using the ground state algorithm.
The cusp in the $^1\Sigma_u$ curve is due to a curve crossing between the $^1\Sigma_u^+$ and $^1\Sigma_u^-$
curves near $r=1.5$ \AA.
}\label{fig:c2}
\end{figure*}

The energies of these twelve states are shown in Table~\ref{table:c2_v5z} and Fig.~\ref{fig:c2}.
In addition the excitation energies of the eight
lowest-lying excited states, as shown in Table~\ref{c2_exc_energies}. These excitation energies
have a mean absolute deviation of 0.02 eV relative to the experimental values.

\begin{table}[h]
\begin{tabular}{c|c|c|c}
\hline
\hline
& & \multicolumn{2}{c}{Excitation energy (eV)}\\\cline{3-4}
State & $R_{\rm eq}/$ \AA\ & Calculated & Experimental \\ 
\hline
$X^1\Sigma_g^+$ & 1.24253 &  0 & 0\\
$a^3\Pi_u$ & 1.312 &  0.07 & 0.09\\
$b^3\Sigma_g^-$ & 1.369 &  0.78 & 0.80\\
$A^1\Pi_u$ & 1.318 &  1.03 & 1.04\\
$c^3\Sigma_u^+$ & 1.208 &  1.16 & 1.13\\
$B^1\Delta_g$ & 1.385 &  1.49 & 1.50\\
$B'^1\Sigma_g^+$ & 1.377 &  1.90 & 1.91\\
$d^3\Pi_g$ & 1.266 &  2.50 & 2.48\\
$C^1\Pi_g$ & 1.255 &  4.29 & 4.25\\
\hline
\hline

\end{tabular}
\caption{
Excitation energies of various states of the carbon dimer, calculated with the cc-pV5Z basis set with a frozen core.
As in previous DMRG studies~\cite{Sharma2015},
the bond length for the $^1\Sigma_g^+$ ground state was taken from FCIQMC in the cc-pVQZ basis~\cite{Booth2011},
while the other bond lengths were taken chosen to be the experimental values~\cite{martin1992c2}.
Errors in the calculated excitation energies (relative to Full CI) are smaller than 1 meV.
The difference between the calculated and experimental excitation energies could be due to
basis set incompleteness, core correlation, relativistic effects, or the interpretation of
the experimental data.
\label{c2_exc_energies}}
\end{table}

\subsection{Tetracene: complete active space calculations}
Singlet fission is a promising phenomenon that could enable more efficient solar energy conversion.
A photon excites a singlet ground state to a singlet excited state, which quickly ``fissions'' into two
lower-energy triplet excited states, thus enabling a single photon
to excite two electrons.
The acenes and their derivatives appear to be among the most promising candidates to harness this phenomenon.

As a final application in this paper,
we performed complete active space (CAS) calculations in three ways:
with a (18e, 18o) pi-orbital active space and either the STO-3G or DZ basis,
and with a (18e, 36o) active space containing a double-pi-orbital manifold in the DZ basis.
All calculations were performed at the ground state singlet geometry optimized at the UB3LYP/6-31G(d)
level of theory.

\begin{table}\label{tab:tetracene}
\begin{center}
\adjincludegraphics[width=0.5\textwidth, clip]{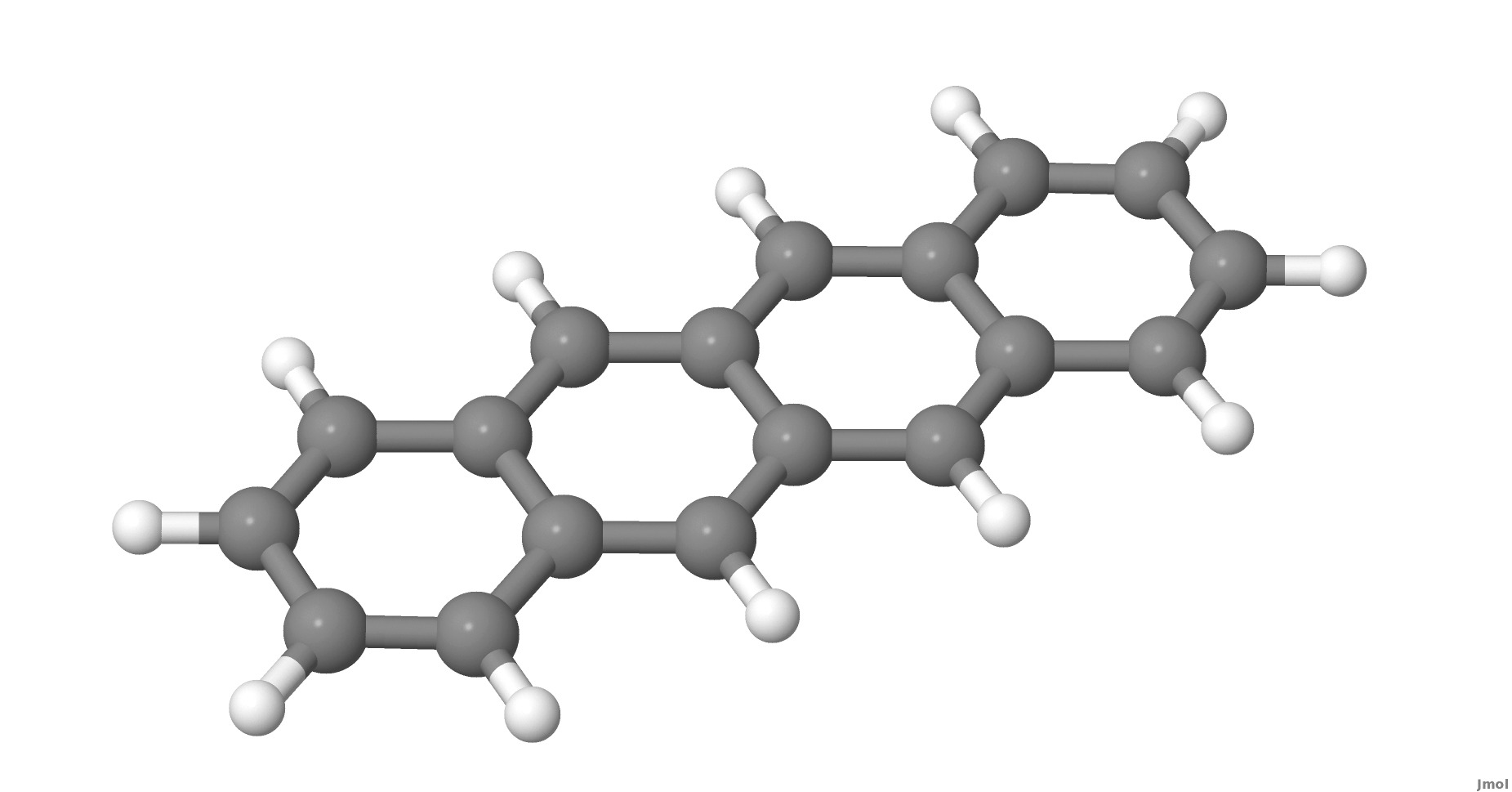}
\end{center}
\begin{tabular}{c|c|c|c|c}
\hline
\hline
& & \multicolumn{3}{c}{Excitation energy (eV)}\\\cline{3-5}
Basis & Active space & $^3B_{2u}$ & $^1B_{2u}$ & $2^1A_g$\\
\hline
STO-3G & (18e, 18o) & 1.867(3) & 6.34(1) &  4.444(3) \\
DZ & (18e, 18o)     & 1.8346(4) & 4.215(2) &  4.3182(4) \\
DZ & (18e, 36o)     & 1.67(2) & 4.37(2) & 4.01(3) \\
\hline
\hline
\end{tabular}
\caption{
Tetracene excitation energies (in eV) relative to the $1^1A_g$ ground state,
calculated using different basis sets and active spaces.
In each case, the pi electrons were chosen to be active, and either a singly- (18o) or
doubly-complete (36o) set of active pi orbitals were used.
Each energy was obtained by an extrapolation with respect to $\Delta E_2$ using five
different $\epsilon_1$ values. A quadratic fit was used to extrapolate to the
CAS limit ($\Delta E_2=0$).
The geometry of the tetracene molecule is shown above the table.
This geometry was optimized for the ground (singlet) state, at the UB3LYP/6-31G(d) level of theory,
by Hachmann et al~\cite{Hachmann2006}.}
\end{table}

In each basis set/active space, we performed three types of runs:
one targeting the lowest two $^1A_g$ states,
one targeting the lowest $^3B_{2u}$ state, and one targeting the lowest $^1B_{2u}$ state.
For each of the three types of runs, natural orbitals were obtained from an HCI run with
$\epsilon_1=5\times10^{-5}$ Ha. Then, runs were performed using $\epsilon_1$ ranging from
$2\times 10^{-4}$ to $2\times 10^{-5}$ Ha, and linear extrapolation to the CASCI limit
was performed with respect to $\Delta E_2$.

Interestingly, as shown in Table~\ref{tab:tetracene}, of the three CASCI calculations, only the (18e, 18o) DZ calculation
produced the correct ordering of the singlet excited state energies.
We believe that dynamical correlation
outside of the active space must be included at least perturbatively to obtain more accurate excitation energies.
This was also the case for pentacene, as reported by Kurashige and Yanai~\cite{kurashige2014theoretical}.
We are currently exploring various methods of including dynamical correlation, including both
contracted and uncontracted multireference perturbation theories with HCI as an active space solver.

\section{Conclusions and Outlook}\label{conclusion}
We have presented an efficient excited-state method using Heat-bath Configuration Interaction and a method for extrapolating the resulting energies to the FCI limit.
We incorporated symmetries including time-reversal symmetry and angular momentum conservation, enabling us to target excited states in different symmetry sectors.
We then used the method to explore the lowest singlet and triplet states in tetracene (relevant to singlet fission)
and to calculate twelve low-lying potential energy surfaces of the carbon dimer in the large cc-pV5Z basis.

We are exploring including one more symmetry: the analog of time-reversal symmetry for angular momentum
(to separately target $\Sigma^+$ and $\Sigma^-$ states). For challenging problems,
we are also extending our extrapolation procedure to the case where the
variational and perturbative steps have different active space sizes, resulting in an extrapolation to the limit of an uncontracted
multireference perturbation theory with a complete active space (CAS) reference, rather than to the Full CI limit.

\begin{acknowledgements}
The calculations in this paper were performed using the University of Colorado's Research Computing cluster. SS and AAH were supported by the startup package from the University of Colorado.
The research was also supported in part by NSF grant ACI-1534965.
\end{acknowledgements}

\bibliographystyle{jchemphys}
\bibliography{hci,c2,singletfission}

\end{document}